\documentclass{JHEP3}

\input{epsf}
\usepackage{epsfig}
\def\G#1#2 { {\displaystyle \Gamma{ #1 \brack #2  } }  }

\def\nn               {  \nonumber  }

\def\period     { \, . }
\def\comma      { \, , }

\def\eqb         {  \begin{eqnarray}  }
\def\eqe           {  \end{eqnarray}  }
\def\nn               {  \nonumber  }

\preprint{MIT-CTP-3330 \\
          UTHEP-466}

 \title{On time-dependent $AdS/CFT$}
 \author{Yuji Satoh \\ Institute of Physics \\
University of Tsukuba \\
Tsukuba, Ibaraki 305-8571 \\ Japan 
\\ \email{ysatoh@het.ph.tsukuba.ac.jp}}
\author{ 
Jan Troost\\      Center for Theoretical Physics \\  MIT \\
    77 Mass Ave \\ Cambridge, MA 02139 \\ USA \\
\email{troost@mit.edu}
   }

 \abstract{We clarify aspects of 
the holographic AdS/CFT correspondence
that are typical of Lorentzian signature, to lay the foundation for a 
treatment of time-dependent gravity and conformal field theory phenomena. 
We provide a derivation of bulk-to-boundary
propagators associated to advanced, retarded and Feynman bulk propagators, and
provide a better understanding of the boundary conditions satisfied by the bulk
fields at the horizon. We interpret the
subleading behavior of the wavefunctions in terms of specific vacuum 
expectation values, and compute
two-point functions in our framework. We connect our bulk methods  
to the closed time path
formalism in the boundary field theory.}

\begin{document}

\section{Introduction}
The AdS/CFT correspondence \cite{Maldacena:1997re} originated 
in part from the comparison of absorption amplitudes calculated
using IIB supergravity, and from the worldvolume action of $D3$-branes \cite{Klebanov:1997kc} in Lorentzian signature
(see e.g. \cite{Klebanov:tk} for a brief review). A clear Euclidean
computational prescription for the duality was formulated
in  \cite{Gubser:1998bc}\cite{Witten:1998qj}. Since then, the AdS/CFT correspondence has been mostly taken into
a Euclidean setting, with great success.

Nevertheless, it has been stressed many times that important physical problems, that may have a natural resolution
in a holographic framework are time-dependent. 
If we want to study processes in gravity such as the formation or evaporation of a
black hole, and we want to obtain a holographic description of the process in terms of a dual conformal
field theory, then time-dependence will surely enter the game.
Clearly, it will not be an easy task to obtain such a description. 

Time-dependent
quantum field theory is a difficult subject by itself, independent of the complications associated to holography
(see  \cite{Schwinger:1960qe}\cite{Bakshi:1963bn}\cite{Bakshi:1962dv}\cite{Keldysh:ud} and references thereto). 
Indeed, like in a classical wave problem, we can (instead of only asking questions about
the overlap of asymptotic states) try to evolve a quantum field theoretical system according to the wave equation of the
quantum field theory at hand, naturally using retarded and advanced propagators (instead of Feynman propagators only). 
These problems are often
harder to solve  than the problems in scattering theory although recently 
good progress has been made thanks to 
numerical methods. One might hope that a better understanding of time-dependent AdS/CFT can shed some light
on time-dependent problems in strongly coupled (supersymmetric conformal) field theories at large N.

In this paper we take one more
 step towards a holographic dictionary between Lo\-rentz\-ian AdS and a time-dependent conformal 
quantum field theory on the boundary. We make a connection between previous works on Lorentzian AdS/CFT,
and we tie together the prescriptions given in the literature 
\cite{Gubser:1998bc}\cite{Balasubramanian:1998sn}\cite{Balasubramanian:1998de}\cite{Balasubramanian:1999ri}\cite{Giddings:1999qu}\cite{Son:2002sd},
  in an elementary approach that we hope clarifies the basic issues. We will assume throughout that  we work in a string theoretic context, where the AdS/CFT correspondence has been tested most convincingly.

We start in section \ref{wave} by briefly reviewing the wave function for a scalar field in $AdS_{d+1}$, and
as an extra, we give an intuitive explanation of the fact that boundary conditions on the scalar field are needed for
low values of the mass. In section \ref{poincare} we review the standard quantisation of a scalar field
in Poincare coordinates. By reinterpreting the bulk Green functions in section \ref{reinterpreting}, we are 
able to relate the regularisation prescription in Minkowksi space to a radial regularisation prescription.
Thus, we link causal properties in the boundary to radial boundary conditions in the bulk.
In section \ref{correlation}, we apply the
formalism to compute two-point functions and to interpret the subleading behavior of bulk solutions
to the equations of motion. We assembled conclusions and remarks in section \ref{conclusions}.

\section{Wave functions on the  Poincare patch}
\label{wave}
In this section, we clarify some  assumptions that underlie the standard quantisation of scalar fields in $AdS_{d+1}$.
That will clear the ground for quantising scalar fields on $AdS_{d+1}$ and deriving their Green functions in the
next section.
\subsection{A rough analysis}
The metric in Poincare coordinates  $(u,x^{\mu})$ is ($u \in {[}0,\infty)$) (see appendix \ref{conventions} for our
conventions) :
\begin{eqnarray}
ds^2 &=& u^{-2}(dx^{\mu})^2 + \frac{du^2}{u^2}, 
\end{eqnarray}
and we will also frequently make use of the radial coordinate $r=\frac{1}{u}$.
To get a first idea of the spectrum for a scalar field, we briefly discuss
solutions to the wave-equation:
\begin{eqnarray}
(\Box-m^2) \Phi(r,x^{\mu}) &=& 0.
\end{eqnarray}
Using separation of variables, we can write a generic solution as 
a linear combination of factorised wave functions
$\Phi(r,x^{\mu})=e^{-i E t} e^{i p_j x^j} R(r)$.
We distinguish two types of solutions, depending on the causal nature
of the Minkowski momentum $p^{\mu}$.
When the Minkowski momentum is spacelike ($p^2>0$), we have:
\begin{eqnarray}
\Phi &=& c e^{-i E t} 
e^{i p_j x^j} r^{-\frac{d}{2}} K_{\nu} (\sqrt{p^2} \frac{1}{r}).
\end{eqnarray}
The solution is regular in the interior (near $r=0$), and (delta-function) normalizable,
for  $0 \leq \nu= \sqrt{m^2+\frac{d^2}{4}}<1$. 
We assume that  the Breitenlohner-Freedman
\cite{Breitenlohner:jf}\cite{Mezincescu:ev} bound $m^2 \ge - \frac{d^2}{4}$ is
satisfied\footnote{
As an aside, we note that an excitation that violates the BF bound is the
analogue of a tachyon in $AdS$. These excitations play a crucial role for strings on 
$AdS_3$, where in the twisted, winding sector, there are stable excitations
 built on the
component of the particle Hilbert space that is tachyonic and associated to the radial
momentum modes \cite{Maldacena:2000hw}. Note also that in the 
euclidean $AdS_3$ setting, the only available modes
are exactly the ones that correspond to these tachyonic modes in the Lorentzian $AdS_3$
(i.e. it is only the continuous representations of $Sl(2)$ that are present for $SL(2,C)$ 
\cite{Gawedzki:1991yu}\cite{Maldacena:2001km}). 
It would be interesting to understand whether one can make sense of these
tachyonic modes in QFT on $AdS$ in general, in an interacting setting.}.
For  timelike momentum ($p^2<0$) the two solutions regular in the interior are:
\begin{eqnarray}
\Phi^{\pm} &=&  c_{\pm} e^{-i E t} e^{i p_j x^j} 
r^{-\frac{d}{2}} J_{\pm \nu} (\sqrt{-p^2} \frac{1}{r})
\end{eqnarray}
(for $\nu$ not an integer).
For 
 $\nu \geq 1$ only $\Phi^+$ is normalizable, while for $\nu<1$ both modes are.
The meaning of this fact will be clarified in the following subsections.

To obtain more insight in the radial behavior of these modes, and in particular,
the way their normalisability properties depend on the value of $\nu$, we introduce
in the next subsection
an auxiliary one-dimensional radial differential equation that will be of good use
throughout our paper.

\subsection{Boundary conditions}
To clarify which eigenfunctions we should consider and what boundary conditions should
be imposed, it is useful to reduce the scalar wave equation to a one-dimensional problem.
In the $(u,x^{\mu})$ Poincare coordinate system, the factors of the function
$\Phi(u,x^{\mu})= \phi(x^{\mu}) u^{d/2-1/2} f(u)$ will satisfy the wave equation in 
Minkowski space with
(mass)${}^2=\lambda$ and the one-dimensional Sturm-Liouville equation
in the radial direction $u$ with eigenvalue $\lambda$ \cite{Bertola:1999vq}:
\begin{eqnarray}
(\Box_M -\lambda) \phi(x^{\mu}) &=& 0 \nonumber \\ 
-{f(u)}'' + \frac{m^2+d^2/4-1/4}{u^2} f(u) &=& \lambda f(u).
\end{eqnarray}
More generally, we  will find it useful to discuss the
inhomogeneous Sturm-Liouville problem
with source $f_s(u)$:
\begin{eqnarray}
{{f(u)}}'' + (\lambda-q(u)) f(u) &=& f_s(u),
\end{eqnarray}
with parameter $
\nu = + \sqrt{m^2+d^2/4} $
and potential $
q(u) = \frac{\nu^2-1/4}{u^2}
$.

\subsection*{Necessity of boundary conditions}

The radial equation for the scalar wavefunction can be written
in terms of the differential operator 
$-\partial_u^2+ \frac{\nu^2-1/4}{u^2}$. Before we discuss this
operator in detail, we consider the problem of wave solutions on the real
half-line ${[} 0, \infty {[}$ to gain some intuition on when it is necessary
to impose boundary conditions on the wavefunction at $u=0$. 

First of all, if we study the problem without potential ($q(u)=0$), then,
if we don't specify a boundary condition
at the end of the half-line ($u=0$)
 for the plane wave (which is thought of as an
eigenfunction of $-\partial_t^2+\partial_u^2$),
 we don't have a well-defined physical problem, and that is
reflected in the fact that the operator $-\partial_u^2$ is not
self-adjoint on the space of functions without specific boundary condition.
We need to specify a boundary condition at zero, because the wave reaches
zero easily, and we need to know how it bounces back to determine its
full evolution. It turns out that the specification of a 
boundary condition is in one-to-one correspondence to the specification
of a self-adjoint extension of the operator $\partial_u^2$ on the 
half-line \cite{RS}.

In general, whether we need to specify a boundary condition at zero,
in a Schr\"odinger problem on the half-line, depends on how fast the potential
grows at zero. If it grows very fast, it takes care of the boundary condition
automatically. The only normalisable solution will have a fixed phase as
it bounces off the wall at $u=0$. If the potential does not grow fast enough, we need
to specify the boundary condition that will pick one out of two normalisable
solutions to the wave equation.

How fast is fast enough ? The critical behavior of the potential is
$\frac{c}{u^2}$. It turns out
\cite{RS} that when $c \ge \frac{3}{4}$, the potential
grows fast enough to have only one normalisable solution. When
$c < \frac{3}{4}$, we need to specify a boundary condition. Since
$\nu^2 = c+\frac{1}{4}$, we find that when $0 \le \nu < 1$, we will need
to specify a boundary condition at zero for the Schr\"odinger problem on the
half line. When $\nu \ge 1$ we don't need to specify a boundary condition,
and there will be only one normalisable solution. 

That explains the pattern we observed above for the normalisability of the scalar
wave functions in $AdS_{d+1}$.
It also shows that we need to impose an extra boundary condition in the
case $0 \le \nu <1$. The choice of boundary condition makes the propagation of fields
on the $AdS$ space well-defined for that mass range. If we do not specify
the boundary condition, we don't know how the scalar wave bounces back
off the boundary of the spacetime.
Thus, we conclude from this analysis that we can just consider the 
eigenfunctions proportional to $J_{\nu}$ when $\nu \ge 1$. However, when
$0 \le \nu<1$, we need to specify boundary conditions. That will single out
a linear combination of $J_{\pm \nu}$, and a single bound state wave
function proportional to $K_{\nu}$ \cite{Bertola:1999vq}\cite{T}. With this extra understanding,
we can systematically treat the quantisation of the $AdS$ scalar field.
Our analysis here can be read as an intuitive restating of facts laid
bare in \cite{Breitenlohner:jf}\cite{Mezincescu:ev}\cite{Bertola:1999vq}.

In this paper we concentrate on the case $\nu \ge 1$\footnote{We moreover will
brush over important subtleties associated to integer $\nu$.}.
For $\nu < 1$ there are two types of conformal boundary 
conditions \cite{Witten:2001ua} that single out the $\Phi^{\pm}$ modes and that exclude
a bound state solution.   
    We hope to 
return to the renormalisation group flow
physics associated to more general boundary 
conditions (see e.g. \cite{Witten:2001ua}\cite{Klebanov:1999tb})
and the interpretation of the bound state
elsewhere. We thus have clarified a little the assumptions that underlie the
ordinary quantisation of scalar fields in $AdS_{d+1}$ and can now proceed with
a clear conscience in standard fashion.

\section{Bulk propagator}
\label{poincare}
In this section we take a first look at the computation of the bulk Feynman,
retarded and advanced propagators (see also \cite{Danielsson:1998wt}\cite{Ryang:1999xm}).
We concentrate on quantizing the modes $\Phi \propto J_{\nu}$ proportional
to the Bessel function with positive index $\nu \ge 1$.
\subsection{Quantisation}
We write the general solution in terms of positive and 
negative frequency components, a normalisation constant $c(p)$ and
annihilation operators $f(p)$:
\begin{eqnarray}
\Phi &=& \int_{E>\sqrt{p_j p^j}} d E \int d^{d-1} p_j f(p) 
c(p) e^{i p.x} r^{-\frac{d}{2}} 
 J_{\nu} (\sqrt{-p^2} \frac{1}{r}) \nonumber \\ &+&
f(p)^{\dagger} c(p)^{\ast} e^{-i p.x}  r^{-\frac{d}{2}} 
 J_{\nu} (\sqrt{-p^2} \frac{1}{r}).
\end{eqnarray}
To normalize the operators $f$, we compute the scalar product between 
the positive frequency eigenfunctions:
\begin{eqnarray}
(\Phi^{(+)} (p),\Phi^{(+)} (p')) &=& -i \int dr d x^j\sqrt{-g} g^{00}
\Phi^{(+)} (p)^{\ast} \partial_{0}^{\leftrightarrow} \Phi^{(+)} (p')
\nonumber \\
&=& (2 \pi)^{d-1} \delta(p_j-p_j') 
c(p)^{\ast} c(p') \int_{0}^{\infty} dr. r^{d-1}.r^{-2}.r^{-d} 
(E+E') \nonumber \\ & &  J_{\nu} ( \sqrt{-p^2} \frac{1}{r})
 J_{\nu}  (\sqrt{-(p')^2} \frac{1}{r}).   \\
&=&  (2 \pi)^{d-1} \delta(p_j-p_j') c(E,p_j)^{\ast} c(E',p_j) (E+E')
\nonumber \\ & &  \int_{0}^{\infty} du u  J_{\nu} ( \sqrt{-p^2} u)
 J_{\nu}  (\sqrt{-(p')^2} u).
\end{eqnarray}
We  have for $\nu>1$:
\begin{eqnarray}
f(s) &=& \int_0^{\infty} s^{1/2} J_{\nu} (su) u 
         \int_0^{\infty} t^{1/2} J_{\nu} (tu) f(t) du dt
\end{eqnarray}
from which it follows that:
\begin{eqnarray}
\int_0^{\infty} du u J_{\nu}(su) J_{\nu} (tu) &=& s^{-1} \delta(s-t).
\end{eqnarray}
Using this formula, we find that the inner product evaluates to:
\begin{eqnarray}
(\Phi^{(+)} (E,p_j),\Phi^{(+)} (E',p_j')) &=& |c(p)|^2 2 (2\pi)^{d-1} \delta(p_j-p_j') 
\delta(E-E').
\end{eqnarray}
Thus we put $c(p)=\frac{1}{\sqrt{2}} (2 \pi)^{-\frac{d-1}{2}}$.
We then postulate:
\begin{eqnarray}
{[} f(p), f^{\dagger}(p'){]} &=& \delta(p-p')
\end{eqnarray}
to quantise the fields. It can then be checked that the quantum fields satisfy the
standard commutation relations recalled in appendix \ref{conventions}.
\subsection{Green functions}
Consider first the Wightman function $G^+$, which we can rewrite in terms of the
Wightman function $G^+_M$ in Minkowski space \cite{Bertola:1999vq}:
\begin{eqnarray}
G^{+}(x,x') &=& \langle 0| \Phi(x) \Phi(x') |0 \rangle \nonumber \\
&=& \int_{\sqrt{p_j p^j}}^{\infty} dE \int \frac{d^{d-1} p_j}{(2 \pi)^{d-1}}
 e^{i p.(x-x')} r^{-\frac{d}{2}}  {r'}^{-\frac{d}{2}} \frac{1}{2} J_{\nu} (\sqrt{-p^2} \frac{1}{r})
J_{\nu} (\sqrt{-p^2} \frac{1}{r'}) \nonumber \\
 &=& \int_0^{\infty} d \lambda \int \frac{d^d p}{(2 \pi)^{d-1}}
e^{i p.(x-x')} \theta(E) \delta(p^2+\lambda) 
 r^{-\frac{d}{2}}  {r'}^{-\frac{d}{2}} \frac{1}{2} J_{\nu}
 (\sqrt{\lambda} \frac{1}{r})
J_{\nu} (\sqrt{\lambda} \frac{1}{r'}) \nonumber \\
           &=& \int_0^{\infty} d \lambda G^+_M(x^{\mu},{x'}^{\mu}; 
\sqrt{\lambda})
 r^{-\frac{d}{2}}  {r'}^{-\frac{d}{2}} \frac{1}{2} J_{\nu} (\sqrt{\lambda} \frac{1}{r})
J_{\nu} (\sqrt{\lambda} \frac{1}{r'}).
\end{eqnarray}
In fact, an analogous analysis makes clear that there is a similar relation between all
Green functions for the Poincare patch of $AdS_{d+1}$ and the Green functions on 
$M^d$. Indeed, the Wightman functions $G^{\pm}$ for a scalar field in $AdS_{d+1}$ and 
those in Minkowski space, $G^{\pm}_M$, are related as above, and the time-ordering 
($\theta$-functions) agree in the Poincare coordinates and Minkowski space. Thus, all
Green functions in Poincare coordinates can be straightforwardly derived from the Green
functions in Minkowski space.

Let us be a little more specific. We will be interested
in computing the retarded, advanced and Feynman bulk Green functions
$G_B \in {\{} G_R,G_A,-G_F {\}} $. We have: 
\begin{eqnarray}
G^{\pm}(x,x') 
&=& \int_0^{\infty} d \lambda
 \int \frac{ d^{d-1}p}{(2 \pi)^{(d-1)}} \int_{C^{\pm}} 
\frac{dE}{2 \pi i} \frac{1}{p^2+\lambda} e^{ip.(x-x')}
 r^{-\frac{d}{2}}  {r'}^{-\frac{d}{2}} \frac{1}{2}
 J_{\nu} (\sqrt{\lambda} \frac{1}{r})
J_{\nu} (\sqrt{\lambda} \frac{1}{r'}). \nn \\ 
 && 
\end{eqnarray}
where the contours $C^{\pm}$ coincide with the Minkowski contours $C^{\pm}$ encircling one of
two poles, as  discussed in appendix
\ref{conventions}. There we also listed
how to obtain the standard  contours $C^B$ for the bulk propagators $G_B$ which can be
expressed as:
\begin{eqnarray}
G_B (x,x') &=& \int_0^{\infty} d \lambda
 \int \frac{ d^{d-1}p}{(2 \pi)^{(d-1)}} \int_{C^B} 
\frac{dE}{2 \pi } \frac{1}{p^2+\lambda} e^{ip.(x-x')}
\nonumber \\ & &  
r^{-\frac{d}{2}}  {r'}^{-\frac{d}{2}} \frac{1}{2}
 J_{\nu} (\sqrt{\lambda} \frac{1}{r})
J_{\nu} (\sqrt{\lambda} \frac{1}{r'}). \label{bulkgreen} 
\end{eqnarray}
The contours $C^B$ enumerated in the appendix
correspond to the usual $\epsilon$-prescriptions:
for the Feynman propagator $G_F:$ $p^2 \rightarrow p^2-i \epsilon$, for the
retarded one $G_{R}:$ $p^2 \rightarrow p^2 - i \epsilon E$ 
and $G_{A}:$  $p^2 \rightarrow p^2 + i \epsilon E$ for the advanced
propagator. That defines the
bulk propagators in $AdS_{d+1}$. 

Up till now, the treatment has been fairly standard. We can compare these 
bulk propagators with the ones obtained in global coordinates, and we
expect them to agree (up to time-ordering) on the grounds that the vacua in Poincare coordinates
and in global coordinates only differ in their vacuum energy, not in
the definition of positive and negative frequency modes \cite{Danielsson:1998wt}\cite{Spradlin:1999bn}. 
We check their agreement
in appendix \ref{comparison}. In the next section,
we will reinterpret the expression for the bulk propagator in our
holographic context.

\section{Reinterpreting the bulk Green function}
\label{reinterpreting}
For our purposes it will be important to be able to take a different perspective
on the bulk Green function $G_B$.
 We have several ways of reading formula (\ref{bulkgreen}) for $G_B$. One is as the Fourier transform
of the Green function for the radial problem (later denoted $g(u,u')$), where the
radial Green function is written in the form of an integral over eigenfunctions.
The second way is to read it as a Bessel transform of a Green function in
Minkowski space. I.e. we can interpret the role of the factor $\frac{1}{p^2+\lambda}$
in two ways, either as associated to a Minkowski Green function, or as 
associated to a radial Green function. That also implies that, 
after implementing the $\epsilon$-prescription, we can either associate
it to the integral over the energy $E$ (leading to standard Minkowski propagators), or
as regularising our radial Green function. To understand the second
perspective better, we collect some useful ingredients.

\subsection{A relation of the bulk propagator to a 1d propagator}
Suppose we have a bulk Green function $G_B$ that satisfies:
\begin{eqnarray}
(\Box_x-m^2) G_B(x,x') &=& -\frac{1}{\sqrt{-g}} \delta(x-x').
\end{eqnarray}
When we naively write it as:
\begin{eqnarray}
G_B &=& \frac{1}{(2 \pi)^d}\int dE \int dp_j e^{-iE(t-t')} e^{i p_j (x^j-{x'}^j)}
g(u,u'), 
\end{eqnarray}
then it is easy to derive that $g(u,u')$ has to satisfy ($\lambda=E^2-p_jp^j$):
\begin{eqnarray}
(\lambda u^2 + u^{d+1} \partial_u(u^{-d+1} \partial_u) -m^2) g(u,u')
&=& - u^{d+1} \delta(u-u').
\end{eqnarray}
When we define a Green function for the 1d problem by: 
\begin{eqnarray}
\partial_u^2 k_{\lambda}(u,u') + (\frac{-m^2+1/4-d^2/4}{u^2} + \lambda) k_{\lambda}(u,u')
&=& - \delta(u,u')
\end{eqnarray}
then $g$ is related to $k_{\lambda}$ by:
\begin{eqnarray}
g(u,u') &=& u^{\frac{d-1}{2}} {u'}^{\frac{d-1}{2}} k_{\lambda}(u,u').
\end{eqnarray}
Thus, we obtain a bulk propagator $G_B$ as the Fourier transform of a Green function
for the radial problem (when we take into account some extra powers of the radial
coordinate). That is a first, quick way to view the abovementioned change of
perspective.

\subsection{Radial bulk propagator}
Next, we discuss how to obtain the radial bulk propagator $k_{\lambda}(u,u')$
 on the interval ${[}u_0,\infty{[}$ (see section 4.10 of \cite{T}).
We introduced an infrared cut-off in the radial direction, $u_0>0$, which excises the
region near the boundary of the $AdS$ space at $u=0$.
We need to study the problem for imaginary values of $\lambda$,
and in particular we will take $\lambda$ to have (at least a small)
positive imaginary part. (This is necessary to make the wavefunction
near the radial origin $r=0$, or $u=\infty$, strictly square integrable.)
After some computation, we find that the radial bulk propagator with 
boundary condition $k(u,u_0)=0$ is given by:
\begin{eqnarray}
k_{\lambda=s^2}(u,u') &=& -\frac{\pi}{2 H_{\nu}^{(1)}(u_0 s) }
u^{1/2} H_{\nu}^{(1)}(us) (u')^{1/2}  (J_{\nu}(u's) Y_{\nu}(u_0 s)-Y_{\nu}(u's)J_{\nu}(u_0 s)) 
\nonumber \\
\quad \mbox{for} \quad u \ge u', 
\label{radprop}
\end{eqnarray}
and a similar expression with $u$ and $u'$ interchanged for $u < u'$.
It can be checked that if we take a naive $u_0 \rightarrow 0$ limit in this expression,
we recuperate the radial Green function for the Sturm-Liouville problem on the half-line \cite{T},
which is directly related to the radial part of the  bulk Green function (\ref{bulkgreen}).

\subsection{Radial bulk-to-boundary propagator}

For later use, we discuss how to obtain a bulk-to-boundary propagator 
(in momentum space) from
the bulk propagator for the one-dimensional problem. 
It is easy to see that the only eigenfunction that behaves nicely
at $u=\infty$, that satisfies the inhomogeneous boundary 
condition $f(u_0)=f_0$ and that is associated to $\lambda=s^2$ (with real and
imaginary part of $s$ larger than zero) is given by:
\begin{eqnarray}
f(u) &=& f_0 \frac{u^{1/2} H_{\nu}^{(1)}(us)}{u_0^{1/2} H_{\nu}^{(1)}(u_0 s)}.
\end{eqnarray}
How do we obtain this solution from the bulk propagator for this
problem ?
In this regularised problem, we can easily derive from the radial propagator
 (\ref{radprop}), for $u' \rightarrow u_0$: 
\begin{eqnarray}
\partial_{u'} k_\lambda(u,u') & \rightarrow &  \frac{ u^{1/2} H_{\nu}^{(1)}(us) }{(u_0)^{1/2} H_{\nu}^{(1)}(u_0 s)},
\end{eqnarray}
yielding the correct (trivial)  bulk-to-boundary 
Green function we found above, as expected from general techniques used
 in solving wave equations (see e.g.
\cite{MF}).

It is  clear that a naive limit of the bulk Green function, where we consider $u_0 \rightarrow 0$, or more
precisely $u_0 s \rightarrow 0$, yields the bulk Green function that we find in the
problem on the real positive half-line \cite{T}. Note though that we 
would neglect terms
that could
 become of equal strength when $us \rightarrow 0$. 
That would cause 
problems in finding the correct  bulk-to-boundary propagator in the 
non-regularised problem. Physically speaking, the high momentum modes on the brane
ruin the limit (-- this is a UV phenomenon --), 
or, in other words, the bulk long distance IR phenomena associated to
the boundary at $u=0$
are not properly taken into account and need to be IR regulated from the bulk supergravity
point of view.  This UV/IR correspondence is familiar by now \cite{Susskind:1998dq}.

Note that the sign of the 
imaginary part of $s=\sqrt{\lambda}$ was decisive in choosing
between the Hankel functions $H^{(1,2)}$. Indeed, had we taken the case
where $\lambda$ has negative imaginary part, we would have obtained very
similar results with $H^{(2)}$ replacing $H^{(1)}$. 
The $\epsilon$-prescription,
or imaginary part of the eigenvalue, determines the radial part of the 
wavefunction uniquely.

%(XXX
%\subsection{One dimensional treatment}
%For purposes of taking the boundary limit of the Green functions that we
%study later, it is useful to note a few facts about the general solution
%to the (inhomogeneous) one-dimensional problem in terms of Green functions.
%The Green function, and the general solution to the inhomogeneous 
%Sturm-Liouville problem is given by:
%\begin{eqnarray}
%G(u,u';\lambda) &=& \frac{\pi}{2 i} u^{1/2} (u')^{1/2} H^{(1)}_{\nu} (us) J_{\nu} (u's) 
%\quad \mbox{for} \quad u' \le u  \nonumber \\
%G(u,u';\lambda) &=& \frac{\pi}{2 i} u^{1/2} (u')^{1/2}  J_{\nu} (us) H^{(1)}_{\nu} (u's) 
%\quad \mbox{for} \quad u' >   u \nonumber \\
%f(u,\lambda) &=& \int_0^{\infty} du' G(u,u',\lambda) f_0(u') 
%\end{eqnarray}
%One can check that the discontinuity of the derivative of $G$ is precisely 
%$1$ at $u$ equal to $u'$ and that $f(u,\lambda)$ is the solution to the inhomogeneous
%Sturm-Liouville problem.
%XXX)

\section{Deep throat}
\label{deepthroat}
We have now assembled the ingredients to be able to derive the bulk-to-boundary
propagators $K_B$ associated to bulk propagators $G_B$, when we cut off the $AdS_{d+1}$ space
at finite radius $r_0 = \frac{1}{u_0}$. We need to reinterpret our bulk propagator as the
Fourier transform of a radial Green function, and take the appropriate regularisation 
prescription into account in the radial part. The Feynman bulk-to-boundary
propagator for instance is the Fourier transform of the one-dimensional bulk-to-boundary propagator
associated to the $p^2 \rightarrow p^2-i \epsilon$ or 
$s^2 \rightarrow s^2+ i \epsilon$  prescription: 
\begin{eqnarray}
K_F(u, x; {x'}) &=& -\partial_u' G_F(u,u') \quad \mbox{at} \quad u'=u_0 \nonumber \\
    &=& 
 \int \frac{ d^{d}p}{(2 \pi)^{d}} 
  e^{ip.(x-x')}  \frac{u^{d/2} H_{\nu}^{(1)}(u \sqrt{-p^2+i\epsilon})}{u_0^{d/2} 
H_{\nu}^{(1)}(u_0 \sqrt{-p^2+i\epsilon})},
 \end{eqnarray}
where we reinstalled the appropriate power of the radial coordinate.
{From} this, it is easy to see that $K_F(u_0)= \delta(x^{\mu}-{x'}^{\mu})$, as expected.
For the advanced and retarded bulk-to-boundary Green functions, we need to adapt our
radial regularisation to the sign of the energy $E$. We obtain: 
\begin{eqnarray}
K_{R,A}(u,x;x') &=&  \partial_u' G_{R,A}(u,u') \quad \mbox{at} \quad u'=u_0 \nonumber \\
        &=& 
 \int \frac{ d^{d}p}{(2 \pi)^{(d)}} 
  e^{ip.(x-x')} ( \theta(\pm E) \frac{u^{d/2} H_{\nu}^{(1)}(u \sqrt{-p^2+i\epsilon})}{u_0^{d/2} H_{\nu}^{(1)}(u_0 \sqrt{-p^2+i\epsilon})}
\nonumber \\
& & 
+ \theta(\mp E) \frac{u^{d/2} H_{\nu}^{(2)}(u \sqrt{-p^2-i\epsilon})}{u_0^{d/2} 
H_{\nu}^{(2)}(u_0 \sqrt{-p^2-i\epsilon})}).
\end{eqnarray}
In this case too, it is easy to see that $K_B(u_0)= \delta(x^{\mu}-{x'}^{\mu})$.

It is clear that the Feynman bulk-to-boundary
Green function is a specific analytic continuation of the Euclidean bulk-to-boundary propagator
as described in \cite{Gubser:1998bc}\cite{Balasubramanian:1998de}.
The bulk-to-boundary propagator contains waves that move inward for positive energy modes,
and outward for negative energy modes. (We consider incoming modes to be the ones that
travel towards the horizon at $r=0$.) We showed that the $\epsilon$-prescription in the Minkowski
part of the bulk propagator determines the 
radial part for the propagator. 

Our derivation illuminates the presciption given in
\cite{Son:2002sd}. Indeed, for the retarded and advanced propagators, we {\em derived} the correct 
boundary conditions at the horizon at $u=\infty$. That is a crucial ingredient
in making the AdS/CFT prescription work. We therefore not only reproduced the fact that indeed
the retarded propagator is associated to incoming boundary conditions on the bulk scalar field
\cite{Son:2002sd}, but we will also see  that the contribution to the two-point
function at the horizon automatically vanishes once the appropriate $\epsilon$-prescription is
taken into account. This puts two ingredients of the prescription in \cite{Son:2002sd} on a firm footing.

%Y ? Connect statement to absorption cross sections, perhaps in section on two-point functions.

\section{Correlation functions}
\label{correlation}
\subsection{Boundary behavior}
We reformulate the boundary value problem for $AdS_{d+1}$ in order
to clarify the structure of the general solution to the equations
of motion. After this general discussion, we return to the
specific conclusions we can draw for the particular choices of bulk-to-boundary
propagators that we defined in the previous section.

{From} the equations of motion, it becomes clear that the most general
(non-normalisable) solution behaves as 
 (see e.g. \cite{Balasubramanian:1998de}\cite{Klebanov:1999tb}):
\begin{eqnarray}
\Phi(u,x^{\mu}) &=& u^{d-\Delta_+} (\phi_-(x^{\mu}) + O(u^2))
+ u^{\Delta_+} (\phi_+(x^{\mu}) + O(u^2))
\end{eqnarray}
where $\Delta_{\pm} = \frac{d}{2} \pm \nu$ and $\nu \ge 0$. It is clear then
that the first term dominates near the boundary at $u=0$ and 
is associated to non-normalisable behavior near the boundary for
$\nu \geq 1$. The second term correponds to leading normalisable boundary
behavior.

%Let's now formulate our boundary value problem in the following way:
%we want to determine the solution to the equation of motion with 
%boundary behavior specified by $\phi_{\pm}(x^{\mu})$. How do we
%write that solution (which may be unique), given the bulk Green
%functions ?
%%
%\subsection{Behavior of bulk Green function}
%XXX(
%We may attempt to write the solution (partly) in terms of boundary-to-bulk
%Green functions derived from bulk Green functions. When we try to do precisely
%that, we discover a surprising feature. When we take the limit of the
%bulk Green function $G_B(+)$ built out of the normalisable ($+$-modes),
%we find that $G(x^{\mu},x')$ behaves as $u^{\Delta_+}$. That is perhaps
%expected, as it is the Green function associated to normalisable modes.
%
%However, when we now use that boundary-to-bulk Green function to write
%a solution to the equation of motion, the solution will be non-normalisable !
%That is because $G_B(x^{\mu}, x')$ for $u' \rightarrow 0$ behaves like
%$(u')^{d-\Delta_+} \delta((x^{\mu})^2)$.
%
%Using the bulk Green function associated to 
%normalisable modes to derive a bulk-to-boundary propagator, we obtain
%a non-normalisable solution to the equations of motion.
%
%)XXX

We note that a generic solution with non-normalisable boundary behavior
specified by $\phi_-$ can  be obtained as
follows: derive a bulk-to-boundary Green function $K_B$ from the bulk
propagator $G_B$,
and write a solution to the equation
of motion satisfying the boundary
condition $\phi_-$. If we do not add any other term to the solution,
 we have also fixed  the subleading $\phi_+$ behavior.
Picking a specific bulk propagator $G_B$ and, hence, $K_B$
can thus be understood as associating a particular subleading behavior
$\phi_+$ to each leading behavior
$\phi_-$\cite{Balasubramanian:1998de}. 
We will analyse and interpret their precise relation in the next section.
 A general solution can be obtained by adding a normalisable
solution to the one obtained in the above prescription. Of course,
adding a normalisable mode will influence the subleading
$\phi_+$ boundary behavior \cite{Balasubramanian:1998de}.

%We can start with $K_F$, and determine the subleading behavior $\phi_+$ in terms
%of the boundary condition $\phi_-$:
%\begin{eqnarray}
%K_F &=&
% \int \frac{ dp^{d}}{(2 \pi)^{(d)}} 
%  e^{ip.(x-x')}  \frac{u^{d/2} H_{\nu}^{(1)}(u \sqrt{-p^2+i \epsilon})}{u_0^{d/2} 
%H_{\nu}^{(1)}(u_0 \sqrt{-p^2+i \epsilon})}
%\nonumber \\
%\Phi(u,x^{\mu}) &=& \int d {x^{\mu}}' \phi_-({x^{\mu}}') K_F(x^{\mu},{x^{\mu}}';u).  
%\end{eqnarray}
%Clearly, the boundary condition is satisfied since $\Phi(u_0,x^{\mu})=\phi_-(x^{\mu}$.
%XXX Leading and subleading behavior by expansion in $u$ ? XXX

\subsection{Subleading behavior}
As a preliminary to studying the two-point functions, we analyse the subleading
behavior of specific wave functions, associated to the bulk-to-boundary
propagators that we derived. First of all, let's study the bulk solution 
associated to the Feynman bulk-to-boundary propagator $K_F$:
%
%a generic value for the second coordinate). It is straightforward
%to obtain:
%\begin{eqnarray}
%z & \rightarrow & \frac{1}{2u} (u'+ \frac{x^2}{u'} ) \nonumber \\
%i G_F(x,x') & \rightarrow & 2^{-\nu-d/2-1} \pi^{-d/2}
%\frac{\Gamma(\nu+d/2)}{\Gamma(\nu+1)} z^{-\nu-d/2} . 1 \nonumber \\
%            & \rightarrow & \frac{1}{2 \nu} \frac{\Gamma(\Delta_+)}{
%\pi^{d/2} \Gamma(\Delta_+-d/2)} u^{\Delta_+} (\frac{u'}{(u')^2+x^2})^{\Delta_+}.
%\end{eqnarray}
%Now, if I follow the Giddings prescription, I need to multiply this by $2 \nu$
%to obtain the bulk-to-boundary propagator. That matches a naive analytic
%continuation of the euclidean bulk-to-boundary propagator (including the
%normalisation), with an extra prescription
%that can be rewritten as $x^2 \rightarrow x^2+i \epsilon$:
\begin{eqnarray}
K_F &=&
 \int \frac{ d^{d}p}{(2 \pi)^{d}} 
  e^{ip.(x-x')}  \frac{u^{d/2} H_{\nu}^{(1)}(u \sqrt{-p^2+i \epsilon})}{u_0^{d/2} 
H_{\nu}^{(1)}(u_0 \sqrt{-p^2+i \epsilon})}
\nonumber \\
\Phi(u,x^{\mu}) &=& \int d {x^{\mu}}' \phi_-({x^{\mu}}') K_F(x^{\mu},{x^{\mu}}';u).  
\end{eqnarray}
Using the expansion of the Hankel functions for small arguments, and the 
Fourier transform of $(-p^2 + i \epsilon)^{\nu}$ \cite{GS}, it is possible to show that the
wavefunction $\Phi$ behaves as:
\begin{eqnarray}
\Phi(u,x^{\mu}) & \simeq & (\frac{u}{u_0})^{d/2-\nu} \phi_-(x ^{\mu})
+ i (\frac{u}{u_0})^{d/2+\nu} \pi^{-d/2} \frac{\Gamma(\Delta_+)}{\Gamma(\nu)}
u_0^{2 \nu} \int d^d y \frac{\phi_-(y)}{((x-y)^2+i \epsilon)^{\Delta_+}},
\quad \ \label{sub}
\end{eqnarray}
from which we easily read off the subleading behavior.
We can do a similar exercise for the retarded and advanced cases, and we find that
the leading and subleading behaviors are: 
\begin{eqnarray}
\Phi(u,x^{\mu}) & \simeq & (\frac{u}{u_0})^{d/2-\nu} \phi_-(x ^{\mu})
+  2 \sin \pi \Delta_+ (\frac{u}{u_0})^{d/2+\nu} \pi^{-d/2} \frac{\Gamma(\Delta_+)}{\Gamma(\nu)}
u_0^{2 \nu} \nonumber \\
& & 
\int d^d y \theta(\pm(x^0-y^0))  \frac{\phi_-(y)}{(-(x-y)_+^2)^{\Delta_+}},
\end{eqnarray}
where we used the Fourier transform of $(p^2 \mp i \epsilon sign(E))^{\nu}$
which can be found in \cite{BP} and we have made use of the generalised function
$x^{\lambda}_+$ defined in \cite{GS} as (roughly) being $x^{\lambda}$ when $x>0$ and
zero otherwise. We have thus explicitly seen how the subleading behavior $\phi_+$
 of the wavefunction
is related to the leading behavior $\phi_-$. The interpretation of the subleading term is
derived in the next subsection.
\subsection{Two-point functions}
To compute two-point functions in the different prescriptions, we follow the appendix of
\cite{Freedman:1998tz} (see also \cite{Muck:1998rr}), 
and write the bulk-to-boundary propagators in momentum space as:
\begin{eqnarray}
K_F (u,p^{\mu}) &=&   \frac{u^{d/2} 
H_{\nu}^{(1)}(u \sqrt{-p^2+i\epsilon})}{u_0^{d/2} H_{\nu}^{(1)}(u_0 \sqrt{-p^2+i\epsilon})}
\nonumber \\
K_{R,A}(u,p^{\mu}) &=& \theta(\pm E) \frac{u^{d/2} 
H_{\nu}^{(1)}(u \sqrt{-p^2+i\epsilon})}{u_0^{d/2} H_{\nu}^{(1)}(u_0 \sqrt{-p^2+i \epsilon})}
+ \theta(\mp E) \frac{u^{d/2} 
H_{\nu}^{(2)}(u \sqrt{-p^2-i\epsilon})}{u_0^{d/2} H_{\nu}^{(2)}(u_0 \sqrt{-p^2-i\epsilon})}.
\end{eqnarray}
We need to get rid of contact terms, and we see  that it is important not to take the limit $u_0 \rightarrow 0$
too fast, not to loose the right normalisation for the two-point function \cite{Freedman:1998tz}.
Following the appendix of \cite{Freedman:1998tz}, we obtain precisely the same boundary
two-point function
in momentum space as the authors of  \cite{Freedman:1998tz} did
in Euclidean signature, but  for the Feynman two-point function we need to replace $p^{2 \nu}$ by
$(p^2-i\epsilon)^{\nu}$ (and there is an overall factor of $i$). 
To obtain this, we crucially made use of the fact that for $u \rightarrow \infty$ the 
bulk-to-boundary Green function vanishes (i.e. $K_B \rightarrow 0$ for $u \rightarrow \infty$)
to (automatically) get rid of a boundary term. We  thus recuperate the two-point
functions in \cite{Son:2002sd}, but in a more transparant fashion.
%\footnote{Useful formulas in this context are:
%$H_{\nu}^{(1)} (z) = \frac{2}{i \pi} e^{-i \nu \pi/2} K_{\nu} (z e^{-i \pi/2}) $ and
%$H_{\nu}^{(2)} (z) = -\frac{2}{i \pi} e^{i \nu \pi/2} K_{\nu} (z e^{i \pi/2}) $.}
We obtain precisely the boundary (Feynman) time-ordered two-point function.
By a familiar Fourier transform, we get the two-point correlator in
position space.

 By plugging in a $\delta-$function
source term for the boundary field following \cite{Balasubramanian:1998de}\cite{Klebanov:1999tb},
 we can interpret the subleading behavior of the 
bulk solution in (\ref{sub}) as the time-ordered expectation value of an operator, in the presence
of another operator insertion up to an overall
normalisation, which we claim coincides with the one found in \cite{Klebanov:1999tb}.
To derive in a precise manner the normalisation constant
requires more care in the regularisation scheme
(see e.g \cite{deHaro:2000xn}\cite{Minces:2001zy}).

We can repeat a similar computation for the retarded and advanced two-point function, but we have
to be careful when defining the two-point function in terms of the action functional. 
After some computation, we find in position space the following action functional for the case
of the retarded bulk-to-boundary propagator 
(where we suppress a double integration
over the boundary space):
\begin{eqnarray}
S {[} \phi_-(x), \phi_-(y) {]} &=& \phi_-(x) \phi_-(y) \theta(x^0-y^0) 
\nonumber \\
& & 
u_0^{2 \nu-d} \frac{\Gamma(\Delta_+)}{\Gamma(\nu)} \pi^{-d/2} 
2 \nu \sin \pi \Delta_+ \cdot (-(x-y)^2)_+^{-\Delta_+}.
\end{eqnarray}
Clearly, a symmetric functional derivative with respect to the two source-terms will not
define an (asymmetric) time-ordered two-point function. The trick we need is familiar from
the closed time-path formalism for quantum field theory (see e.g. \cite{Chou:es} for a 
review). We can formally introduce two independent sources $\phi_-^1(x)$ and $\phi_-^2(y)$
which are time-ordered with respect to each other. The retarded two-point
function is then defined by the functional derivative with respect to these two sources
(see  \cite{Chou:es} for more detail) and becomes:
\begin{eqnarray}
\langle \theta(x^0-y^0) O(x) O(y) \rangle &=&-i  \theta(x^0-y^0) \nonumber \\
& & 
u_0^{2 \nu-d} \frac{\Gamma(\Delta_+)}{\Gamma(\nu)} \pi^{-d/2} 2 \nu 
\sin \pi \Delta_+ \cdot
(-(x-y)^2)_+^{-\Delta_+}.
\end{eqnarray}
Note that $(-(x-y)^2)_+$ assures that the correlator is vanishing
when $(x-y)^\mu$ is spacelike.
Clearly, this formal trick takes into account the physically evident fact that the operator
$O(x)$ (associated to the source $\phi_-(x)$) has the same time-ordering with respect to
$O(y)$ as the source term $\phi_-(x)$ has with respect to $\phi_-(y)$.
The interpretation
of the subleading behavior of the bulk wave function is then in terms of the expectation value
of an operator in the presence of an operator inserted at an earlier time. We can repeat a 
similar story for the advanced propagator.

What we basically observe is that time-ordering in the bulk and on the boundary agree
in Poincare coordinates, and that the bulk $\epsilon$-prescription, via the radial boundary
conditions, is directly reflected in the $\epsilon$- or time-ordering prescription in 
the boundary field theory.

\section{Conclusions}
\label{conclusions}
By treating carefully the Feynman, retarded and advanced bulk propagator,
we noticed they could be rewritten in terms of a Minkowksi propagator
and a radial Green function. Making use of that key fact, we used
the $\epsilon$-prescription associated to the causal structure of the
bulk propagator to regulate the radial part of the bulk wavefunctions
appropriately, which automatically
lead to the vanishing of the boundary term in the
two-point functions associated to the horizon.  
In this way we derived the boundary conditions on the
bulk wavefuntions that are suited for each choice of propagator, thus
putting the prescription used in \cite{Son:2002sd} on a firm footing.
We derived the bulk-to-boundary propagators from the bulk propagators
with some care, and interpreteted the two-point functions and the 
subleading behavior of the bulk solution to the wave equation thus
making the general statements in \cite{Balasubramanian:1998de} more
specific.
We thus clarified key issues.

{From} our analysis it is clear that we obtain a full set of two-point 
functions, as described in the literature on time-dependent quantum field
theory. We already briefly indicated the relevance of the closed time-path formalism
to the interpretation of the two-point functions in section \ref{correlation}. 
For higher-point functions,
we naturally expect to be able to derive correlators associated to 
different time-orderings of the operators and at that point the extensive
formalism \cite{Chou:es} of closed time-path ordering\footnote{We thank Yoav Bergner and
Luis Bettencourt for explaining parts of the formalism to us.} 
will start playing an important role. (See also \cite{Maldacena:2001kr} for the
relevance of the formalism at finite temperature.)
We hope to return to this 
issue in the future.

Once we have a more complete dictionary for time-dependent AdS/CFT, we should
be able to study enigmatic time-dependent processes in gravity (like
the formation and evaporation of a black hole), or, perhaps more realistically for now,
 time-dependent processes in strongly coupled field theories at large
N using gravity. To make the dictionary more complete, it would be useful to
analyze what the interesting and manageable initial conditions are for the 
wave equation, and how they translate from bulk into boundary and vice versa.
We took a step towards making that analysis well-founded.

\subsection*{Note added}
While we were writing up our work, the paper \cite{HS} appeared on the archive,
which follows a different route to try to justify the prescription
of \cite{Son:2002sd}.

\section{Acknowledgements}
Thanks to Yoav Bergner,
Luis Bettencourt, Dan Freedman, Ami Hanany, Pavlos Kazakopoulos, Asad Naqvi, Dam T. Son
and David Tong
for discussions.
 Our
research was supported in part by
University of Tsukuba Research Projects and the U.S. Department of Energy
under cooperative research agreement \# DE-FC02-94ER40818.

\appendix
\section{Conventions}
\label{conventions}
\subsection{Coordinates and metrics}
Our conventions are as follows. The space in which we embed
$AdS_{d+1}$  has metric:
\begin{eqnarray}
ds^2 &=& - (dX^0)^2+ (dX^1)^2 + \dots + (dX^d)^2-(dX^{d+1})^2.
\end{eqnarray}
The space $AdS_{d+1}$ is defined by:
\begin{eqnarray}
-(X^0)^2+ (X^1)^2 + \dots + (X^d)^2 - (X^{d+1})^2 = -l^2
\end{eqnarray}
and we put $l=1$. We will frequently use the nomenclature $AdS_{d+1}$ for
the cover of this space.
Our Poincare coordinates we define as follows\footnote{Note that $r$ is
a logical radial coordinate, increasing as we go towards the boundary, while
$u=\frac{1}{r}$. We have mostly plus signature in both the
embedding and the Minkowski space.}:
\begin{eqnarray}
X^{\mu} &=& \frac{x^{\mu}}{u} \quad \mu \in {\{} 0,1, \dots, d-1 {\}}
\nonumber \\
X^d &=& \frac{1-u^2}{2u}- \frac{x^2}{2u} \nonumber \\
X^{d+1} &=& \frac{1+u^2}{2u}+ \frac{x^2}{2u} \nonumber \\
r &=& \frac{1}{u} \nonumber \\
t &=& x^0 
\end{eqnarray}
and the metric in these coordinates is:
\begin{eqnarray}
ds^2 &=& u^{-2}(dx^{\mu})^2 + \frac{du^2}{u^2} \nonumber \\
ds^2 &=& r^2(dx^{\mu})^2 + \frac{dr^2}{r^2}.
\end{eqnarray}
The index $j$ will run over the spatial coordinates of the Minkowski
space.

\subsection{Green functions in curved spaces}
We discuss our conventions for Green functions in curved space which mostly
coincide with those of \cite{BD}. 

\noindent

$\bullet$ \ The Green functions for a scalar field $\phi(x)$ are defined by:
\eqb
   G^+(x,x') &=& \langle 0 \vert \phi(x) \phi(x') \vert 0 \rangle \nn \\
   G^-(x,x') &=& \langle 0 \vert \phi(x') \phi(x) \vert 0 \rangle \nn \\
  iG(x,x') &=& \langle 0 \vert [\phi(x), \phi(x')] \vert 0 \rangle \nn \\
  G^{(1)}(x,x') &=& 
     \langle 0 \vert \{ \phi(x), \phi(x') \} \vert 0 \rangle  \\
  G_R(x,x') &=& -\theta(t-t') G(x,x') \comma \qquad 
  G_A(x,x') \ = \ +\theta(t'-t) G(x,x') \nn \\
  iG_F(x,x') &=& 
   \langle 0 \vert T \bigl( \phi(x) \phi(x') \bigr) \vert 0 \rangle 
    \ = \ -\frac{i}{2} \Bigl( G_R(x,x') + G_A (x,x') \Bigr) 
    + \frac{1}{2} G^{(1)} (x,x') \nn  
\eqe

\bigskip
Alternatively, we have for the bulk propagators $G_B$:
\begin{eqnarray}
G_R &=& i \theta(t-t') (G^+-G^-) \nonumber \\
G_A &=& -i \theta(t'-t) (G^+-G^-) \nonumber \\
-G_F &=& i \theta(t-t') G^+ + i \theta(t'-t) G^-.
\end{eqnarray}
These bulk Green functions all have the same source term
(including the sign).

\noindent
$\bullet$ \ The differential equation satisfied by $\phi(x)$ is:
\eqb
  ( \Box - m^2 ) \phi(x) &=& 0 \comma \qquad 
    \Box = \frac{1}{\sqrt{-g}} \partial_\mu \sqrt{-g}g^{\mu\nu} \partial_\nu    
  \period
\eqe

We assume the static geometry with the metric
\eqb
   ds^2 & = & g_{00} (x^i) dt^2 + g_{ij}(x^i) dx^i dx^j
    \comma \qquad (i,j = 1,2,...) \period
\eqe

\bigskip

\noindent
$\bullet$ \ The differential equations satisfied by the Green functions are:
\eqb
   && (\Box_{x} -m^2) G^\pm (x,x') =  (\Box_{x} -m^2) G(x,x')
   = (\Box_{x} -m^2) G^{(1)} (x,x') = 0 \nn \\
   && (\Box_{x} -m^2) G_{R,A} (x,x') = \frac{-1}{\sqrt{-g}} \delta(x-x')
      \\
  && (\Box_{x} -m^2) G_{F} (x,x') = \frac{+1}{\sqrt{-g}} \delta(x-x')
   \nn
\eqe

where we made use of the equations:
\eqb
  \Box (AB) &=& (\Box A) B + 2 g^{\mu\nu}\partial_\mu A \partial_\nu B + A (\Box B)
   \nn \\
  \Box \theta (t) &=& \frac{1}{\sqrt{-g}} \partial_0 \sqrt{-g}g^{00} \delta (t)
          \ = \  g^{00} \delta'(t) 
\end{eqnarray}

and the canonical commutation relations are:
\begin{eqnarray}
   {[} \phi(x) , \dot{\phi} (x') {]}_{t=t'} &=& 
     \frac{-i}{g^{00} \sqrt{-g}} \delta^{(d-1)} (\vec{x}-\vec{x'})
   \comma \quad ( \vec{x} = (x^i) ). 
\eqe
\subsection{Green functions in Minkowski space}
Our conventions for Minkowksi space agree with the ones we specified for a 
general curved space. Then the Green functions $G_B$ for Minkowski space 
 and expectation values $G^{\pm}$ are associated
to the following contours:
\begin{eqnarray}
G^+ (x,x';\lambda) &=& \int \frac{d^dp}{(2 \pi)^{(d-1)}} e^{ip.(x-x')}
\theta(E) \delta(p^2+\lambda) \nonumber \\
&=& \int \frac{ d^{d-1}p}{(2 \pi)^{(d-1)}} \int_{C^+} 
\frac{dE}{2 \pi i} \frac{1}{p^2+s^2} e^{ip.(x-x')}
\end{eqnarray}
where $C^+$ is the contour that circles $E_p = \sqrt{p_jp^j + s^2}>0$ 
clockwise.
The same expression is valid for $G^-$ where $C^-$ encircles
$-E_p<0$ counterclockwise.

For the bulk Green functions that we are interested in, and in
our conventions, we obtain:
\begin{eqnarray}
G^R (x,x';\lambda)
&=& i \int \frac{ d^{d-1}p}{(2 \pi)^{(d-1)}} \int_{C^R} 
\frac{dE}{2 \pi i} \frac{1}{p^2+s^2} e^{ip.(x-x')}
\end{eqnarray}
where the retarded contour $C^R$ is just above the real axis, leaving
the poles underneath (and directed toward $+\infty$), 
while for $G^A$ we find the same expression
for the integral, but now with a contour just below the real $E$-axis
(and oriented toward $+\infty$). For $-G_F$ we also have the
same expression, but now the contour $C^F$ dips at $-E_p<0$ and
rises at $E_p>0$. The associated $\epsilon$-prescriptions are 
easily worked out.

\subsection{$G^{\pm}$ in coordinate space}
We can compute the Minkowski Wightman functions if we regularize $x^2$ as $x^2 \rightarrow
x^2 \pm i \epsilon sign(t)$ to make the integral in the last step well-defined:
\begin{eqnarray}
G^{\pm} (x) &=& \int \frac{d^{d-1} p_j}{(2 \pi)^{d-1}} \frac{1}{2 E_p}
e^{-i E_p t} e^{ip_j x^j} \nonumber \\
& =&
\int_0^{\infty} \frac{d p p^{d-2}}{(2 \pi)^{d-1}} \int d\Omega_{d-2} \frac{1}{2 E_p}
e^{-i E_p t} e^{i p x \cos \theta} \nonumber \\
 &=& \frac{1}{(2 \pi)^{d/2+1/1}} \pi x^{-d/2+3/2} \int_0^{\infty} d p p^{d/2-1/2}
e^{-i E_p t}\frac{1}{E_p} J_{d/2-3/2} (px) \nonumber \\
&=& (2 \pi)^{-d/2} (\frac{s}{\sqrt{x^2}})^{d/2-1} K_{d/2-1} (s \sqrt{x^2 \pm i \epsilon sign(t)}).
\end{eqnarray}

\section{Poincare goes global}
\label{comparison}
The purpose of this appendix is to compare the bulk propagators in Poincare coordinates
that we derived in section \ref{poincare}, to the propagators in global
coordinates.
\subsection{Feynman propagator}
Let's concentrate on the Feynman propagator. We will be able to compute it using the
technique discussed in section \ref{poincare}, and we can then compare it to the Feynman propagator in global coordinates.
To compute the Feynman propagator, we regularize with the $\epsilon$ prescription 
$\frac{1}{\lambda+p^2-i\epsilon}$ in momentum space. The Fourier transform \cite{GS} is then
given by:
\begin{eqnarray}
i G_F &=&   (2 \pi)^{-\frac{d}{2}} (\sqrt{x^2+i \epsilon})^{\frac{-d+2}{2}} (r r')^{-\frac{d}{2}}
\int_0^{\infty} \frac{d \lambda}{2} \lambda^{\frac{d-2}{4}}
J_{\nu}(\sqrt{\lambda} u) J_{\nu}(\sqrt{\lambda} u') K_{\frac{d-2}{2}}
(\sqrt{\lambda} \sqrt{x^2+i \epsilon}) \nonumber \\
 &=& (2 \pi)^{-\frac{d+1}{2}} e^{-i \pi \frac{d-1}{2}}
((z+i\epsilon)^2-1)^{-\frac{d-1}{4}} Q_{\nu-\frac{1}{2}}^{\frac{d-1}{2}} (z+i\epsilon)
\end{eqnarray}
where
$z=\sigma+1= \frac{1}{2} (\frac{r}{r'} + \frac{r'}{r}+r r' x^2) $
and we used that $0<r<\infty$.
We moreover made use of the formula:
\begin{eqnarray}
\int_0^{\infty} x^{\mu+1/2} J_{\nu}(\beta x) K_{\mu}(\alpha x) J_{\nu} (xy)
(xy)^{1/2} dx &=& (2 \pi)^{-1/2} \alpha^{\mu} \beta^{-\mu-1} y^{-\mu-1/2}
e^{-(\mu+1/2)\pi i} \nonumber \\ 
& &  (z^2-1)^{-\mu/2-1/4} Q^{\mu+1/2}_{\nu-1/2} (z)
\end{eqnarray}
which holds for
\begin{eqnarray}
y>0; Re(\alpha)> |Im(\beta)|; Re(\nu)>-1; Re(\mu+\nu)>-1; 
2 \beta y z &=& \alpha^2 + \beta^2+ y^2.
\end{eqnarray}
These conditions are satisfied once the $\epsilon$ regularisation prescription is taken into
account. Now, using the connection between the Legendre function $Q$ and the hypergeometric
function \cite{AS} (for $|z|>1$ and otherwise by analytic continuation):
\begin{eqnarray}
Q^{\mu}_{\rho}(z) &=& e^{i \mu \pi} 2^{-\rho-1} \sqrt{\pi}
\frac{\Gamma(\rho+\mu+1)}{\Gamma(\rho+3/2)} z^{-\rho-\mu-1}
(z^2-1)^{\mu/2} \nonumber \\
& &  F(1+\rho/2+\mu/2,1/2+\rho/2+\mu/2; \rho+3/2; z^{-2})
\end{eqnarray}
we are able to rewrite the Feynman (i.e. causal) Green function as:
\begin{eqnarray}
i G_F &=&  2^{-\frac{d}{2}-\nu-1} \pi^{-\frac{d}{2}} 
\frac{\Gamma(\nu+\frac{d}{2})}{\Gamma(\nu+1)}
z^{-\nu-\frac{d}{2}} F(d/4+\nu/2+1/2, d/4+\nu/2; \nu+1; (z+i\epsilon)^{-2}). 
 \nn \\
&&  \label{GFb6}
\end{eqnarray}
This matches with the Feynman Green function computed in 
\cite{Burgess:ti} in global coordinates.
We thus verified explicitly that the Feynman propagator agrees in 
Poincare and global coordinates. That is as expected since it is known that
the vacuum based on the concept of positive frequency modes using
Poincare time, agrees with the vacuum in global coordinates, up to some
constant vacuum energy (see e.g. \cite{Danielsson:1998wt}\cite{Spradlin:1999bn}).
 We therefore expect the Feynman propagator to
agree in both coordinate systems, and they do.

\subsection{Retarded (advanced) propagator}
We briefly discuss the retarded (and advanced) propagator
in the Poincare patch. Since the Wightman functions $G^{\pm}$ will be given by the
prescription $x^2 \pm i \epsilon \mbox{sign} x^0$, as in Minkowski space
(see appendix \ref{conventions} for the latter),
and taking into account the definition of $G_R$ in terms of $G^{\pm}$,
we conclude that the retarded Green function in Poincare coordinates
$G_R$
will be given by: 
\begin{eqnarray}
G_R &=& \theta(t-t') {[} iG(z+ i \epsilon) - iG(z-i \epsilon) {]} 
\end{eqnarray}
as in global coordinates (since the Feyman functions agree),
 except for the leading $\theta$-function, which differs in Poincare and global coordinates. 
Here,  $G(z)$ is given by
$G_F$ in (\ref{GFb6}) without the $i\epsilon$ prescription.
A similar expression can be obtained for $G_A$.

\end{document}